%% file: Magazine.tex
\documentclass[journal]{IEEEtran}
\usepackage{subcaption}
\usepackage{amsmath}
\usepackage{amsfonts}

\usepackage[ruled,vlined]{algorithm2e}

\usepackage{array}

\usepackage{textcomp}
\usepackage{stfloats}
\usepackage{url}
\usepackage{verbatim}

\usepackage[noadjust]{cite}
\usepackage{xcolor}

\usepackage{amsmath}
\usepackage{amssymb}
\usepackage{mathtools}
\usepackage{graphicx}
\usepackage{epstopdf}
\usepackage{lipsum}

\usepackage{cuted}
\usepackage{amsthm}
\usepackage{soul}
\input{notation}

\usepackage{makecell}               % For wrapping text in cells (especially headers)
\usepackage{enumitem,kantlipsum}

\begin{document}

\title{ Indoor Positioning for Public Safety: Role of UAVs, LEOs, and Propagation-Aware Techniques
%\huge Enhancing Public Safety Networks with Next-Gen Positioning\\
}
\author{
~Gaurav~Duggal,~Harish~K.~Dureppagari,~Harpreet~S.~Dhillon,~Jeffrey~H.~Reed and~R.~Michael~Buehrer
\thanks{The authors are with Wireless@VT,  Bradley Department of Electrical and Computer Engineering, Virginia Tech,  Blacksburg,
VA, 24061, USA. Email: \{gduggal, harishkumard, hdhillon, reedjh, rbuehrer,\}@vt.edu.\\ The support of NIST PSCR PSIAP through grant 70NANB22H070, NSF through grants CNS-1923807, CNS-2107276 and NIJ graduate research fellowship through grant 15PNIJ-23-GG-01949-RESS is gratefully acknowledged.\\ }
}

\maketitle

\begin{abstract}
Effective indoor positioning is critical for public safety, enabling first responders to locate at-risk individuals accurately during emergency scenarios. However, traditional Global Navigation Satellite Systems (GNSS) often perform poorly indoors due to poor coverage and non-line-of-sight (NLOS) conditions. Moreover, relying on fixed cellular infrastructure, such as terrestrial networks (TNs), may not be feasible, as indoor signal coverage from a sufficient number of base stations or WiFi access points cannot be guaranteed for accurate positioning. In this paper, we propose a rapidly deployable indoor positioning system (IPS) leveraging mobile anchors, including uncrewed aerial vehicles (UAVs) and Low-Earth-Orbit (LEO) satellites, and discuss the role of GNSS and LEOs in localizing the mobile anchors. Additionally, we discuss the role of sidelink-based positioning, which is introduced in 3rd Generation Partnership Project (3GPP) Release $18$, in enabling public safety systems. By examining outdoor-to-indoor (O2I) signal propagation, particularly diffraction-based approaches, we highlight how propagation-aware positioning methods can outperform conventional strategies that disregard propagation mechanism information. The study highlights how emerging 5G Advanced and Non-Terrestrial Networks (NTN) features offer new avenues to improve positioning in challenging indoor environments, ultimately paving the way for cost-effective and resilient IPS solutions tailored to public safety applications.
\end{abstract}

\section{Introduction}
Past public safety wireless use cases have focused on enhancing communications signal coverage in both outdoor and indoor environments with a focus on forming resilient communication links to ensure connectivity during emergencies. This is because, during emergencies, the infrastructure can be damaged, affecting communication. Another critical aspect of public safety scenarios is the need for indoor positioning, which is crucial in situations such as firefighting and active shooter incidents \cite{li20225g}. Positional information of at-risk individuals, including emergency responders, police, firefighters, and medical personnel involved in the rescue effort, can both ensure their safety and help locate vulnerable people inside buildings, leading to a more effective emergency response. Since mobility across building floors is more challenging for emergency responders than movement within a single floor, it is crucial not only to enhance 3D positioning accuracy but also to ensure an accurate floor estimate, improving z-axis (vertical) positioning precision.
%Conventional positioning systems like GNSS struggle in indoor environments due to weak signal quality, rendering GNSS-based localization unreliable indoors. Other positioning systems with fixed infrastructure, like cellular BSs and WiFi Access Points acting as anchors for positioning purposes, remain unreliable as they might be susceptible to infrastructure damage. In light of these challenges, we employ swiftly deployable mobile anchors, e.g., UAVs or NTN, to enable public safety systems, especially for emergency applications.

%The FirstNet Authority \cite{li20225g} led the creation of a dedicated nationwide broadband network using spectrum reserved for the public safety community (LTE band 14). \textcolor{blue}{Need to elaborate on FirstNet}.

Now, a key question arises: ``How to build a ubiquitous infrastructure that supports a wide range of mobile anchors for the emerging public safety and emergency services?". To promote the development of public safety and emergency services, the National Telecommunications and Information Administration (NTIA) in the United States established the First Responder Network Authority (FirstNet) in 2012~\cite{firstnet2018} to provide a nationwide broadband network with a special focus on public safety agencies and first responders. In support of this initiative, the Federal Communications Commission (FCC) designated LTE band $14$ ($700\,$MHz) for FirstNet, allocating a bandwidth of $20\,$MHz. The establishment of FirstNet subsequently led to the formation of the FirstNet Alliance Group for public safety, which has paved the way for new possibilities, including the development of proximity-based services using long-term-evolution (LTE) and new radio sidelink (NR SL). Proximity Services (ProSe) enable direct device-to-device communication without routing through traditional cellular infrastructure. %\mike{Is this opinion or requirement?} \gd{its a requirement in 3GPP standards} 
Positioning and ranging are fundamental requirements of any ProSe designed for public safety. Given the growing interest in location-based services for public safety and the automotive industry, the 3rd Generation Partnership Project (3GPP) has introduced sidelink-based location services in Release $18$~\cite{9927255}. Recognizing the growing importance of UAVs in public safety and emergency applications, the FCC recently adopted new rules for drone operations in the $5$ GHz band. These regulations ensure secure and reliable UAV communication, enabling UAVs to support emergency response, search-and-rescue missions, and disaster management efforts, making them a key component of rapidly deployable mobile anchors. Furthermore, the integration of NTN into 5G New Radio (NR) presents a transformative opportunity, as NTN can serve as either a complementary positioning infrastructure or a potential GNSS replacement, thereby extending location services to GNSS-denied environments, including disaster scenarios and remote areas~\cite{dureppagari_ntn_10355106,dureppagari2024leo}.
% \par
% NTN is another promising candidate that has the potential to provide a complement positioning infrastructure or a potential replacement to GNSS due to their superior link budget, higher operating bandwidth, and large forthcoming
% constellations~\cite{dureppagari_ntn_10355106,dureppagari2024leo}.
\par
For a viable Indoor Positioning System (IPS) targeted for public safety applications, our goal is to overcome key challenges while balancing trade-offs in logistics, cost, and technical constraints. The first challenge is to ensure reliable signal coverage within buildings. To elaborate, positioning relies on simultaneous signal coverage from multiple anchors at known locations; however, achieving this inside buildings is extremely challenging. For example, in Global Navigation Satellite Systems (GNSS), satellites orbiting at approximately 20,000 km above the Earth's surface serve as anchors. The wireless signals they transmit experience significant attenuation, both due to free-space path loss over the long transmission distance and additional losses as they penetrate building structures, resulting in poor indoor coverage. On the other hand, terrestrial cellular networks are designed to prioritize strong coverage from one or two base stations to minimize inter-cell interference, and this again leads to inadequate indoor signal coverage across multiple anchors.
The second challenge is minimizing the dependence on pre-installed infrastructure, as emergencies may cause power outages or infrastructure damage, resulting in loss of coverage. This rules out WiFi-based positioning since it depends on pre-installed access points. We also have various commercial location services such as those by Google and Apple,  %\cite{ni2022experience},
which integrate GNSS, cellular, WiFi, and motion sensors to offer good indoor positioning performance; however, again their reliance on preinstalled infrastructure makes them unsuitable for public safety applications.
The third challenge arises from Non-Line-Of-Sight (NLoS) conditions in wireless signal propagation in the indoor scenario, resulting in substantial multipath effects within the environment, which, in turn, degrade positioning accuracy. Multipath occurs when a wireless signal interacts with objects in its environment, such as structural elements of a building in our case, including windows, doors, floors, and ceilings. These interactions result in indirect paths that are longer than the Euclidean paths between the transmitter and receiver, which degrades the positioning accuracy.
Finally, to facilitate deployment, ensure interoperability, and minimize costs, we require a system architecture that leverages existing technologies based on 3GPP standards, enabling reliable indoor positioning tailored to public safety needs. To tackle these challenges, this paper presents the following key contributions.
\begin{itemize}[wide, labelindent=0pt]
    \item \textbf{Emerging System Architectures}: We investigate system architectures for rapidly deployable IPS, where mobile anchors with position knowledge are deployed outside buildings with the goal of localizing indoor nodes. Our study examines the role of LEO satellites, UAVs, GNSS, and sidelink in enabling a practical IPS aligning with the current focus on 3GPP standards.
    \item \textbf{The Impact of Outdoor-to-Indoor (O2I) Signal Propagation}: We examine O2I signal propagation from outdoor mobile anchors to indoor nodes within buildings and identify diffraction from window edges as a key wireless propagation mechanism in this NLoS scenario. These diffraction paths contain essential position information, and utilizing them can enhance positioning accuracy.  
    \item \textbf{New Positioning Approaches}: We present a propagation-aware positioning approache that utilizes propagation mechanism information as a fundamental factor in improving positioning accuracy. This is achieved through specific geometric positioning methods designed for Transmission, Reflection, and Diffraction propagation mechanisms.
\end{itemize}

% \begin{itemize}
%     \item Motivation: positioning scenarios, FirstNet, E911, 3GPP TR 22.872- positioning use case related to firefighter safety
%     \item Implementation opportunities: Both infrastructural with NTN/Sidelink and architectural in 3GPP. 
%     \item Modeling Outdoor-to-Indoor signal Propagation 
%     \item Previous NLoS positioning techniques:
%     Geometric techniques \cite{guvenc2009survey, zekavat2011handbook}, Non-Geometric techniques - fingerprinting, pattern matching/statistical methods/ AI/ML direct and assisted and \textbf{Hybrid}.     
% \end{itemize}

\begin{figure}[!t]
\centering
\includegraphics[clip,width=1\linewidth]{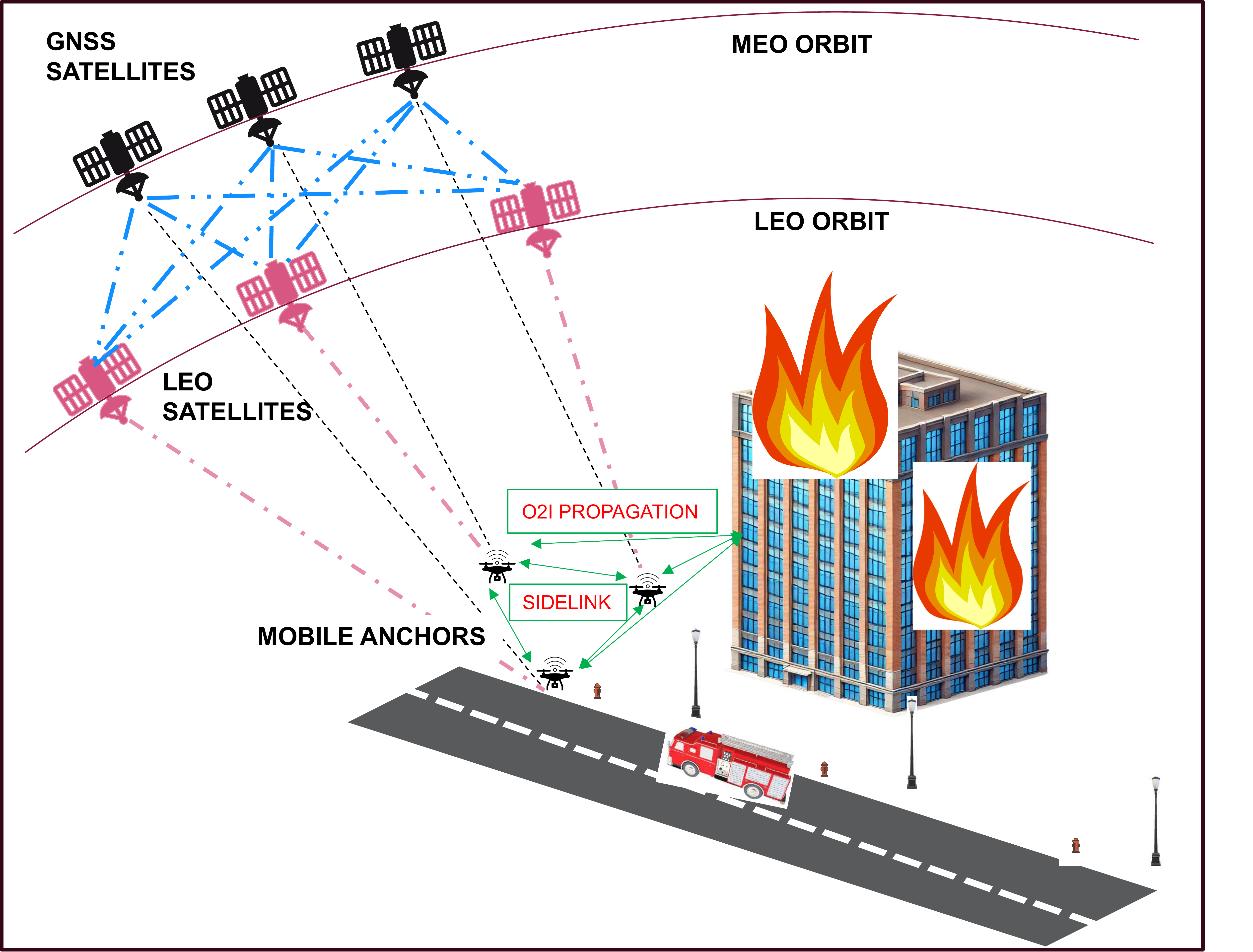}
    \caption{IPS consisting of mobile anchors that determine their position based on various combinations of LEO/GNSS satellites. The mobile anchors can establish wireless links amongst themselves and with indoor nodes, leveraging the 3GPP standard for sidelink. We also leverage insights about the O2I signal propagation to develop new positioning techniques.}
    \captionsetup{font==small}
    \label{fig_positioning_system}
    \vspace{-15pt}
\end{figure}

\section{System Architectures}
IPS generally relies on anchors with known positions to locate indoor nodes. In emergency situations, rapidly deploying such anchors near buildings can greatly enhance indoor signal coverage while keeping costs low. These anchors might be UAVs or temporary fixed equipment installed close to the building. However, accurately determining the position of these mobile anchors is critical for IPS to function effectively. %In the following sections, we detail the evolution of 3GPP standards that pave the way for rapidly deployable mobile anchors, which are essential for accurately locating indoor nodes in challenging environments.
In the following sections, we explore the evolution of 3GPP standards that facilitate the deployment of mobile anchors for IPS. We begin by examining the role of UAVs as mobile anchors and their integration into 3GPP cellular networks. Next, we introduce sidelink positioning (SLP), a promising feature introduced in 3GPP Release 18, which enables device-to-device (D2D) positioning independent of the core network. We further discuss how UAVs can serve as enablers for SLP, forming a cooperative localization framework that enhances positioning accuracy in public safety scenarios. Finally, we explore the feasibility of LEO-based NTN positioning, assessing its potential as a mobile anchor localization solution in GNSS-denied environments and emergency response scenarios. Additionally, we discuss how LEO satellites themselves can serve as mobile anchors, extending coverage even for indoor environments. 

% Alternatively, we could also use LEO-satellites as anchors. These satellites operate closer to the earth as compared to GNSS (between 160 km - 900 km) and could offer better indoor coverage.   
\subsection{UAVs as Mobile Anchors}

UAV standards in 3GPP have advanced significantly since Release 15, which introduced LTE-based aerial UE connectivity studies.~\cite{9927255}. The primary challenges identified included interference mitigation, UAV identification, and mobility enhancements. These were addressed through improvements in interference detection techniques, multi-cell measurement reporting, and UAV-specific handover mechanisms. Release $17$ expanded on these foundations, focusing on network-assisted UAV command and control (C2) communications, authorization procedures via Uncrewed Aerial Systems (UAS) Traffic Management (UTM), and remote UAV identification. With the advent of $5$G, cellular connectivity for UAVs has moved beyond traditional ground-based infrastructure, incorporating features such as beamforming and dynamic spectrum sharing to optimize aerial communication. 

In Release $18$, $3$GPP has further refined UAV integration into 5G-Advanced by introducing height-dependent measurement reporting, flight path reporting, and subscription-based UAV identification to ensure regulatory compliance. Enhanced radio resource planning mechanisms, such as Detect and Avoid (DAA) and Broadcasting UAV ID (BUID), enable safer UAV operations by allowing authorities to track and manage aerial traffic effectively. Additionally, RAN$1$ is investigating UAV-specific uplink beamforming in Frequency Range $1$ (FR$1$) to minimize interference. Looking ahead to Release $19$, advancements will likely include further optimizations for UAV-specific conditional handovers, improved mobility and interference management, and AI-driven resource allocation strategies, bridging the path toward a fully autonomous UAV ecosystem within 6G networks.

Having said that, the important question is how to obtain the location information of the mobile anchors. The accurate localization of mobile anchors is crucial as their estimated positions serve as reference points for determining the locations of target UEs, particularly in challenging indoor environments. A conventional approach is to equip mobile anchors with GNSS receivers and leverage GNSS-based positioning. However, GNSS performance is severely impacted in dense urban environments due to signal attenuation, multipath interference, and NLoS conditions. The challenge is even more pronounced for mobile anchors deployed near buildings, where GNSS signals may be weak or unavailable. In light of this, the integration of non-terrestrial networks (NTN) into $5$G new radio (NR) has opened up the possibility of developing a new positioning infrastructure using NR signals from LEO satellites. However, leveraging LEO constellations for high-precision positioning is not trivial and necessitates significant enhancements at both the system and algorithmic levels, which we discuss in detail in Section~\ref{sec:nr-ntn}.

\subsection{Sidelink Positioning}
% \begin{itemize}
%     \item Sidelink public safety simulation parameters. 
%     \item Needs less anchors to achieve same performance (Collaboration)
%     \item Collaborative can mitigate effect of NLoS bias
%     \item No additional HW required (backhaul related)
%     \item Sidelink PRS vs Regular PRS- more flexibility in terms of signal
%     \item RTT and TDOA measurements are supported
% \end{itemize}
Another promising approach is leveraging UAVs for SLP, an emerging feature in 3GPP standards aimed at applications such as public safety, vehicular communication, and emergency response systems. Introduced in Release $18$, SLP enables devices to exchange position-related signals (for instance, positioning reference signals, or PRS) directly via sidelink communication, without relying on conventional network infrastructure. This capability is particularly beneficial in GNSS-denied environments or during emergencies when terrestrial networks are inaccessible. UAVs equipped with 5G NR sidelink capabilities can serve as anchors by broadcasting positioning reference signals to nearby UEs, thereby assisting in accurate location estimation for first responders and other public safety applications. Furthermore, SLP not only facilitates positioning measurements between mobile anchors (such as UAVs) and UEs but also enables direct inter-node measurements among UEs, thereby enabling cooperative localization and providing improved accuracy. Cooperative localization is particularly effective when an insufficient number of anchors are visible, e.g., when only two anchors are in view. Additionally, SLP supports various positioning methods, including time-of-arrival (ToA), time-difference-of-arrival (TDoA), round-trip-time (RTT), and angle-of-arrival (AoA). Notably, methods such as RTT do not require precise synchronization between nodes, making them particularly attractive for rapidly deployable positioning systems. These characteristics position SLP as a robust, scalable, and adaptable solution for next-generation public safety systems. Next, we detail sidelink positioning implementation and its integration into public safety systems.

\begin{figure}[t]
     \centering
     \includegraphics[width=0.9\linewidth]{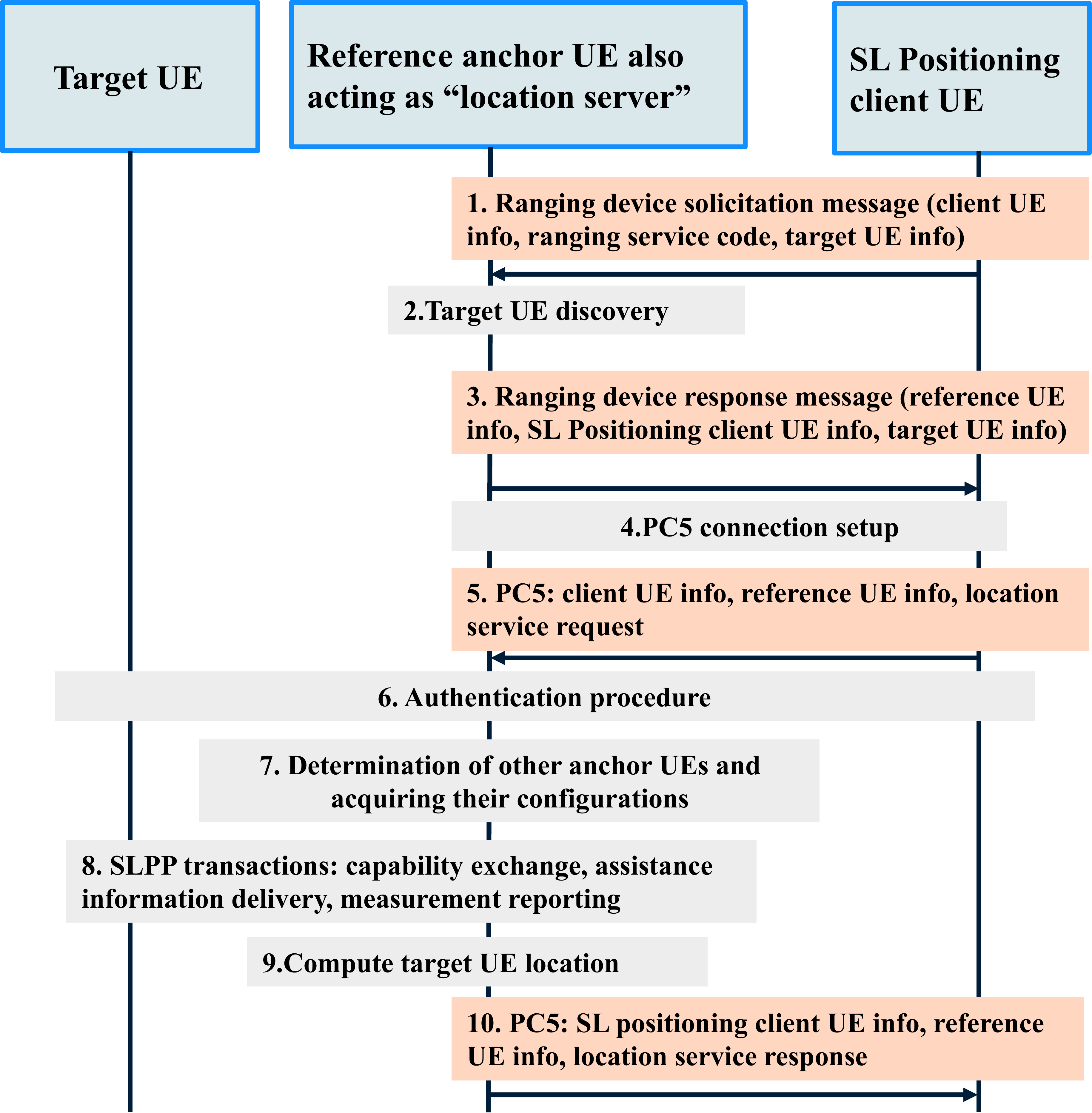}
    \caption{Signal flow of SLP in out-of-coverage scenario.}
    \captionsetup{font==small}
    \label{fig:signal::flow}
    \vspace{-15pt}
\end{figure}

SLP can be considered for three network coverage scenarios for public safety and vehicle-to-everything (V2X) use cases, namely in-coverage, partial coverage, and out-of-coverage, as described in~\cite{3gpp::38845}. %In-coverage scenario refers to a case where two UEs connected by sidelink are within network coverage. Partial coverage implies one of the UEs is inside the network coverage while the other UE is outside the network coverage. 
In this paper, we are interested in out-of-coverage scenarios, where UEs that are outside network coverage are connected by sidelink, i.e., SLP can operate without any network or GNSS coverage. In Fig.~\ref{fig:signal::flow}, we illustrate a high-level overview of the SLP procedure in an out-of-coverage scenario. Putting this signal flow into the context of mobile-anchor aided indoor localization for public safety, illustrated in Fig.~\ref{fig_positioning_system}, the SLP client UE can be a first responder device that may be human-operated. Note that this is an example signal flow for the purpose of illustration; the actual signal flow will likely depend on the implementation. The UE requiring emergency services is the target UE that needs to be localized, and the UAVs equipped with 5G NR sidelink are the anchor UEs. 

In step $1$, the SLP client UE broadcasts a solicitation message to initiate anchor UE to discover the target UE. The reference UE that successfully discovers the target UE responds to the SLP client UE. Then, the SLP client UE selects one or more anchor UEs for the SLP service (step $3$). In step $4$, the reference anchor UE also acts as a location server, and the client UE sets up a PC5 connection. Following this, in step $5$, the client UE sends a location service request to the LMF. Following this, in step $6$, an authentication procedure is performed to determine whether the client UE is authorized to determine the SLP service. In step $7$, the LMF may determine additional anchor UEs for positioning. In step $8$, SLPP transactions between the LMF and target UE occur. This includes capability exchange, assistance data delivery, measurement request, and response. Finally, the LMF computes the UE location based on the measurements collected from the target UE. In step $10$, the LMF reports the UE location to the SLP client in response to the request in step $5$.
% \begin{figure}[t]
%     \centering
%     \includegraphics[width=\linewidth]{figs/deployment_scenarios_slp.pdf}
%     \caption{Deployment scenarios for SLP.}
%     \label{fig:deployment::scenario}
% \end{figure}

\subsection{NR-NTN}\label{sec:nr-ntn}
% \textcolor{blue}{Complementary GPS}
% \begin{itemize}
%     \item Enables position knowledge for mobile anchors
%     \item No additional HW to connect to NTN (supported by NR-NTN standards). Its an alternative/complementary to GPS.
%     \item More anchors available hence helps solve poor GDOP with GPS due to mobile anchors being close to the building.
%     \item Better link budget offered by NTN as compared to GPS.
% \end{itemize}
% \begin{figure}[t]
%      \centering
%      \includegraphics[width=\linewidth]{figs/LEO-PNT-TUDelft.pdf}
%     \caption{NTN positioning, navigation, and timing.}
%     \label{fig:ntn_pnt}
% \end{figure}
Starting from Release $17$, NTN was introduced in 5G NR, supporting cellular services through satellites in LEO, medium-earth orbits (MEO), and geostationary-earth orbits (GEO). As NR-NTN continues to evolve, there is an opportunity to provide positioning as an additional service offered by satellite network operators (SNOs) delivering cellular services over NTN. Unlike TNs, NTN systems are not currently mandated to provide location services. However, in the future, SNOs may need to build location services into their NTN offerings to meet regulatory requirements and manage location-based content delivery and subscriptions. Although GNSS will remain an essential component for PNT, it is crucial to understand whether LEO constellations can offer a promising alternative/complement to PNT using $6$G cellular infrastructure, as illustrated in~\figref{fig_positioning_system}. 

LEO satellites, orbiting at around $600$ km, offer superior link budgets due to enhanced signal strength compared to MEO satellites, which orbit at $20,000$ km. Improved signal strength, coupled with larger bandwidth, improves multipath resolution and reduces time-to-first-fix (TTFF). NTN is particularly beneficial in disaster scenarios where terrestrial infrastructure may be unavailable. Additionally, LEO-based systems can eliminate the need for GNSS radios (as of Release $19$, NR-NTN connectivity requires a UE to resolve its location using GNSS), reducing UE power consumption and complexity. With these advantages, beyond localizing mobile anchors, LEO-based NTN can also serve as mobile anchors themselves to provide extended coverage and facilitate indoor positioning. These attributes position NTN as a promising complementary or even alternative positioning solution to GNSS, particularly in scenarios where GNSS coverage is unreliable or insufficient. 

However, compared to GNSS, which is specifically designed for continuous and precise positioning, repurposing communication-focused LEO satellites for positioning poses several challenges, including intermittent signal transmissions, narrow and frequency-separated beams which are less suitable for positioning compared to the broader beams in GNSS, limited phase coherence which hinders coherent combining of signals across time, and high variability in delay and Doppler shifts adds additional complexity in detecting and tracking PRS signals. Addressing these challenges is critical for unlocking the full potential of NTN-based positioning. Readers are recommended to read~\cite{dureppagari_ntn_10355106,dureppagari2024leo} for more details. % In the next section, we outline the key design considerations and necessary enhancements required to enable accurate and reliable positioning using LEO-based NTN.

\section{The Impact of Signal Propagation}
A major challenge for indoor positioning is the complexity of signal propagation in indoor environments. In particular, multipath propagation and the absence of Line-of-Sight (LoS) paths are of concern. Multipath signal propagation can be thought of as propagation along specific ray paths called multipath components (MPCs) that result from the interaction of the wireless signal with objects in the environment through four distinct mechanisms - reflection, diffraction, transmission, and diffuse scattering. This view of propagation comes from a high-frequency approximation for modeling signal propagation along rays and is based on the computational electromagnetic techniques of Geometrical Optics (GO) combined with the Uniform Theory of Diffraction (UTD). This approach not only offers physical insights into signal propagation but also facilitates the simulation of wireless signal propagation across large study areas, something that prior methods based on the numerical computation of Maxwell's equations struggled to achieve. This has led to the development of various raytracing-based GPU accelerated signal propagation simulators to investigate signal propagation \cite{he2018design}.

\subsection{Propagation Mechanisms}
In wireless signal propagation, a surface is considered smooth if the wavelength of the incident wave is much larger than the surface roughness. In this case, part of the energy is reflected, generating a reflection MPC, while another part is transmitted, forming transmission MPCs, with some energy also absorbed. Building walls, modeled as layered dielectrics, contribute to both reflection and transmission MPCs. Diffraction occurs due to edges like those found on windows, doors, and walls, generating diffraction MPCs, while diffuse scattering results from rough surfaces, dispersing energy in all directions.

\subsection{O2I Signal Propagation}
\label{section_o2i_signal_propagation}
% \begin{figure}[!htbp]        \centering
%         \includegraphics[clip, trim=3.5cm 9cm 4.5cm 9cm, width=0.9\linewidth]{figs/FAP_stat_BW_400_MHz_fap_threshold_20dB.eps}
%     \caption{\textcolor{blue}{To Do}}
%     \label{fig_P_FAP}
% \end{figure}

% \begin{figure}[!tbp]        \centering
%         \includegraphics[clip, trim=0cm 0cm 0cm 0cm, width=1\linewidth]{figs/FAP_stat_BW_400_MHz_fap_threshold_20dB.eps}
%     \caption{\textcolor{blue}{To Do}}
%     \label{fig_P_FAP}
% \end{figure}

Previous studies have shown that in indoor-to-indoor scenarios, the primary propagation mechanisms are LoS and reflection paths \cite{amiri2023indoor}. In contrast, Outdoor-to-Indoor (O2I) signal propagation is primarily governed by diffraction paths \cite{duggal2025diffractionaidedwirelesspositioning}. The diffraction MPCs, originating from outdoor anchors, interact with window edges on the same building floor as the node, thereby capturing vertical axis position information. Additionally, whether this O2I signal propagation mechanism extends to signals transmitted from LEO satellites remains an open question.
Now, among all the MPCs existing between a fixed outdoor anchor and an indoor node, if the transmission MPC are sufficiently attenuated, the diffraction MPCs become the next shortest. As a result, these diffraction paths can be isolated using the first-arriving path principle \cite{duggal2025}. 
From \cite{duggal2025}, using a RayTracing campaign, we found out that for FR$2$ ($24$\,GHz\,-\,$48$\,GHz) and FR$3$ ($7$\,GHz\,-\,$24$\,GHz) frequency bands, diffraction paths can be isolated from all the other paths for a majority of indoor node locations using the first-arriving path principle whereas for FR$1$ bands, since transmission paths are not sufficiently attenuated, this principle fails to isolate the diffraction paths.

\section{Positioning in NLoS scenarios}
\begin{figure*}
    \centering
    \begin{subfigure}{0.68\linewidth}  % Adjust width as needed
        \centering
        \includegraphics[clip, , trim=0cm 0cm 0cm 0.05cm, width=1\linewidth]{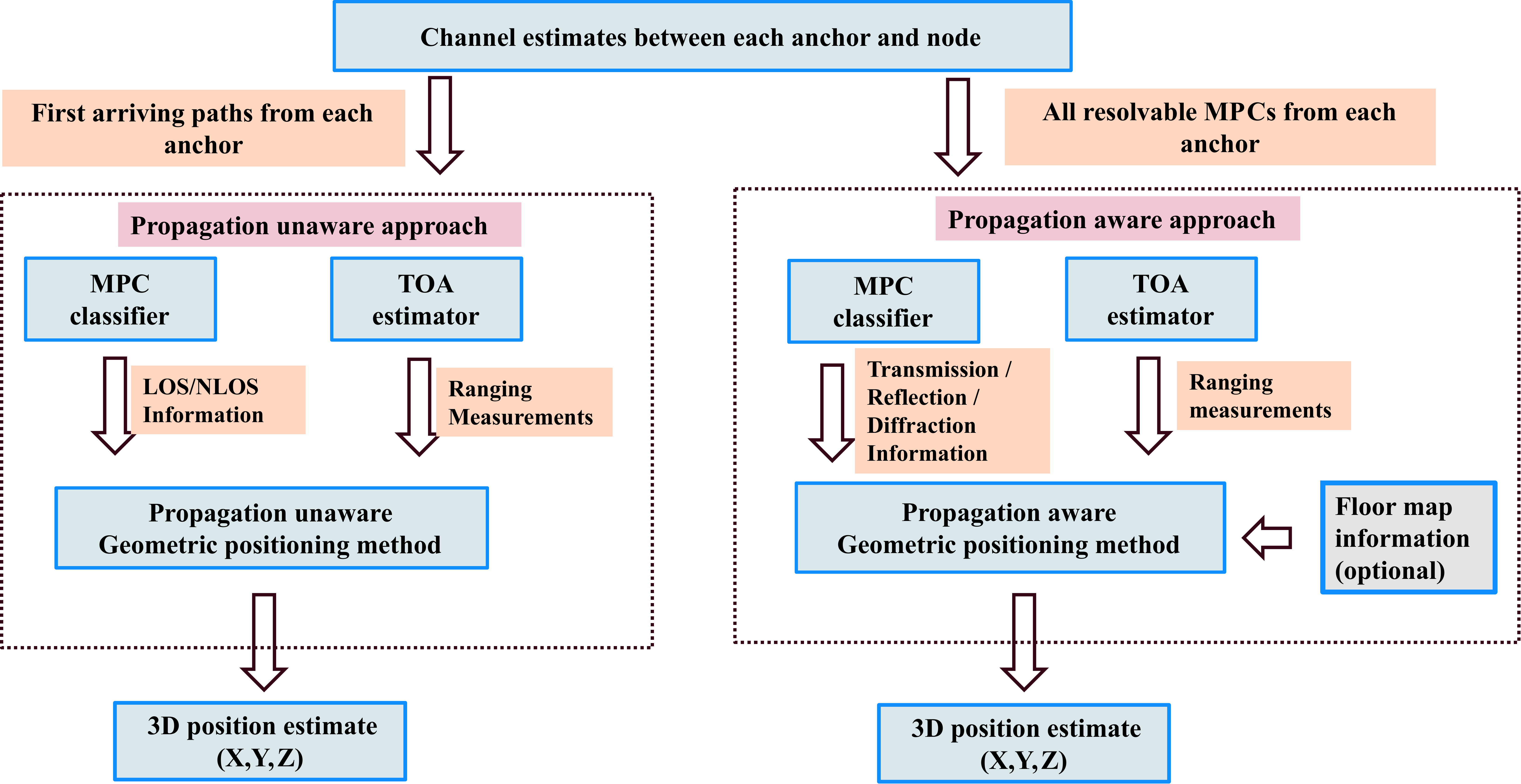}
        \caption{Propagation unaware vs Propagation aware approach to $3$D positioning - Flowchart}
        \label{fig_PU_vs_PA_plot}
    \end{subfigure}
    \hfill
    \begin{subfigure}{0.28\linewidth}
        \centering
        \includegraphics[width=1\linewidth]{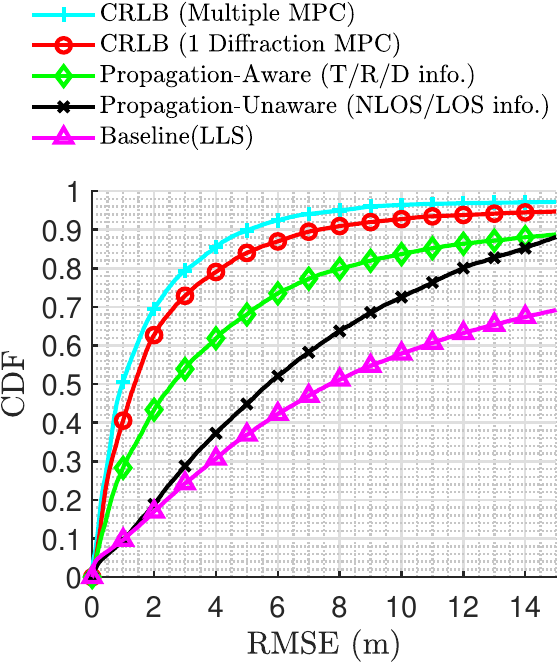}
        \caption{Comparison of Positioning performance}
        \label{fig_positioning_performance}
    \end{subfigure}
    \captionsetup{font==small}
    \caption{The propagation-unaware approach to positioning relies on propagation-unaware geometric positioning methods, which rely on isolating a single MPC per anchor from which we extract LoS/NLoS labels and ranging measurements. In contrast, propagation-aware methods use propagation-aware geometric positioning methods that can leverage several MPCs per anchor. For each MPC we require propagation mechanism information and the ToA measurements. Introducing propagation mechanism information greatly improves the positioning performance.}
    \label{fig:main}
    \vspace{-15pt}
\end{figure*}

% \begin{figure*}[!htbp]        \centering
%         \includegraphics[clip, trim=0cm 0cm 0cm 0cm, width=1\linewidth]{figs/PU_vs_PA_plot.eps}
%     \caption{The propagation unaware approach to positioning relies on
%     propagation unaware geometric positioning methods which rely on  isolating a single MPC per anchor from which we extract LoS/NLoS labels and ranging measurements. In contrast propagation aware methods use propagation aware geometric positioning methods that can leverage several MPC per anchor. For each MPC we require propagation mechanism information and the ToA measurements. Introducing propagation mechanism information greatly improves the positioning performance.}
%     \label{fig_PU_vs_PA_plot}
% \end{figure*}

ToA-based positioning in LoS scenarios is relatively straightforward, whereas NLoS scenarios require a fundamentally different approach. In NLoS environments, we typically try to leverage the position information in multipath
signals. To extract position information from the multipath, we aim to enhance the resolvability of multipath components by having sufficient bandwidth to obtain a multipath channel estimate between each anchor and the node. From the channel estimates, we extract ToA-based ranging measurements. Now NLoS positioning methods, i.e., those which assume at least some ranging measurements come from NLoS conditions, operate on these ranging measurements to produce an estimate of the position. These methods can be broadly classified into two categories based on the information available about the ranging measurements. We have (a) propagation-unaware methods, which require just LoS/NLoS information for each ranging measurement, and (b) propagation-aware methods, which require propagation mechanism information for each ranging measurement.
\subsection{Propagation Unaware Approach}
Observe \figref{fig_PU_vs_PA_plot}, in the propagation unaware approach, we isolate a single path called the First Arriving Path (FAP) \cite{duggal2025} from the multipath channel estimate between each anchor and node by using a threshold based approach. Next, for each FAP between an anchor and node we obtain the associated ToA based ranging measurement and classify if the FAP is LoS or NLoS. Since signals traveling along NLoS paths are generally lower in intensity as compared to signals propagating along direct paths, their classification into LoS and NLoS can be formulated as a hypothesis testing problem \cite{zekavat2019handbook}. Alternatively, machine learning techniques can also be utilized to determine the LoS and NLoS labels. This approach is also investigated in the 3GPP standards in Release $18$ as an application of machine learning, referred to as the Assisted-ML method for determining LoS/NLoS labels \cite{3gpp_ts_38_843_r18}. Now, NLoS paths are generated using multiple interactions with objects in the environment based on the four propagation mechanisms outlined earlier and are typically longer than the Euclidean distance between the corresponding anchor and the node. The excess path length with respect to the Euclidean distance is referred to as the {\em NLoS bias}. This has led to the development of several geometric positioning techniques designed to mitigate the adverse effects of NLoS bias \cite{guvenc2009survey}. These methods depend on accurately identifying LoS and NLoS labels, allowing for separate treatment of LoS and NLoS paths, which ultimately enhances positioning performance. Since these methods do not assume knowledge of the propagation mechanism, they are appropriately named propagation-unaware methods. The Iterative Parallel Projection Algorithm (IPPA) is a computationally efficient propagation-unaware geometric positioning technique, making it well-suited for edge computing applications such as public safety positioning applications. Designed to mitigate NLoS bias, it employs set-theoretic concepts and utilizes information from both LoS and NLoS paths to enhance positioning accuracy.
\subsection{Propagation Aware Approach}
More recently, propagation-aware techniques have emerged, enabling a more refined approach to positioning. Each MPC between the anchor and the node can be further categorized on the basis of the underlying propagation mechanism responsible for its formation. Next, we have propagation-aware geometrical techniques that utilize ToA-based ranging measurements corresponding to each MPC to obtain the final position estimate. The improvements to the positioning results are due to two main reasons. First, knowledge of the underlying propagation mechanism helps to extract more position information as compared to without. Secondly, rather than isolating a single MPC for each anchor-node link, this approach also offers the flexibility to leverage multiple ToA measurements from the same anchor. This effectively increases the number of anchors by introducing new virtual anchors, thus further enhancing positioning accuracy. Next, we illustrate three propagation-aware positioning methods based on mathematical path length models of transmission, reflection, and diffraction. For each propagation mechanism, we can formulate the corresponding propagation-aware geometrical positioning technique as a least squares estimation method with the goal of minimizing the least squares error between the ranging measurements and its corresponding parametric path length model. The minimization is done over candidate node locations.

\subsubsection{Geometric Positioning Method for Transmission and Reflection MPCs}
\begin{figure*}[!t]
\centering
\includegraphics[clip,width=1\linewidth]{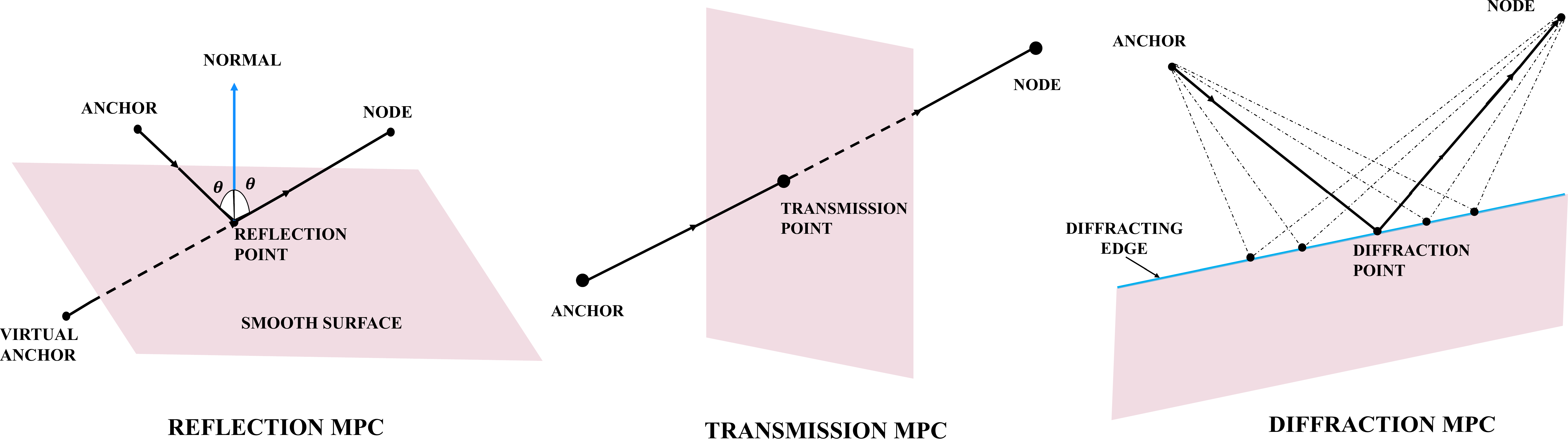}
    \captionsetup{font=small}
    \caption{Propagation-dependent geometric positioning methods are based on various mathematical path models corresponding to different propagation mechanisms. These methods estimate node positions by minimizing the least squares error between the ranging measurements and their respective path models over possible node positions.}
    \label{fig_pa_geometric}
    \vspace{-15pt}
\end{figure*}

Observe in \figref{fig_pa_geometric}, for a fixed anchor-node pair, the transmission path passes through a planar surface obstacle, such as a wall. Refraction through the wall thickness is neglected as it is negligible compared to the overall signal propagation distance. Consequently, the path length is approximated as the Euclidean distance between the anchor and the node. Next, we have reflection paths that result from reflection with planar surfaces. In this case, the path length exceeds the Euclidean distance between the anchor and the node. However, on applying Snell's law of reflection, the anchor can be mirrored across the reflecting surface to create a virtual anchor. Note that this transformation may require {\em a priori} knowledge of the wall's location, obtained from a floor map, or alternatively, additional signal measurements such as the Angle of Arrival (AoA) and Angle of Departure (AoD) to estimate the reflection point. With this transformation, the reflection path length becomes equivalent to the Euclidean distance between the virtual anchor and the node. Since transmission and reflection path lengths can be transformed to the Euclidean distance between the anchor and the node, they can also be linearized by employing a squaring operation \cite{zekavat2019handbook}, thus resulting in a single-shot linear least squares implementation.
\subsubsection{Geometric Positioning Method for Diffraction Paths}
Finally, diffraction takes place when a wireless signal encounters edges, such as windows or building corners. For a specific anchor, edge, and node position, the diffraction point on the edge is uniquely identified by minimizing the sum of two Euclidean distances. The first distance is from the anchor to the diffraction point on the edge, while the second distance is from the diffraction point to the node. This approach is based on Fermat's principle of least time and can be implemented as a numerical optimization method. Alternatively, there is a closed-form expression for locating the diffraction point on the edge \cite{duggal2025diffractionaidedwirelesspositioning}. The diffraction path length is a non-linear function in node coordinates, therefore we use an iterative non-linear least squares estimation method. To determine the indoor position of the node, an intermediate step involves estimating the locations of diffraction points for the diffraction MPC from each anchor. One way to achieve this is to assume precise position knowledge of all possible diffracting edges - in our case, the locations of all window edges for a given building. This information can be obtained from the building floor map and serves as one way to implement the geometric positioning method for diffraction MPCs \cite{duggal20243d}. However, position knowledge of the diffracting edges is not an explicit requirement and can be relaxed using the following observations. First, in the O2I signal propagation scenario, observe that the diffraction MPCs are formed by diffraction with the edges of the windows located on the same building floor as the node. Hence, the vertical coordinate of the diffraction point is approximately the same as the vertical coordinate of the indoor node. Secondly, instead of determining the outdoor anchor coordinates in global coordinates, we can obtain them in local building coordinates. Combining these two observations enables us to relax the requirement of the position of the diffracting edge, which results in a more practical geometric positioning technique for diffraction paths called D-NLS \cite{duggal2025diffractionaidedwirelesspositioning}. Note that the anchor positions in local building coordinates could be obtained by first obtaining the positions in global coordinates and then transforming them to local coordinates with a geo-referenced floor plan of the building. Alternatively, we could leverage the provision in the $3$GPP standards to include measurements from non $3$GPP sensors. %\cite{3gpp_ts_38_305_r18}.
For example, the vertical coordinates of the mobile anchors could be obtained using barometers by calculating the height above the ground. The distance to the building can be obtained using vision-based sensors and the inter-anchor distances computed using RTT measurements over the sidelink protocol.

\subsection{Comparison of the Positioning Approaches}
\label{section_positioning_performance_comparisons}
The plot in \figref{fig_positioning_performance} illustrates the effect of incorporating propagation awareness into our positioning methodology for the public safety system, showing improvements in positioning accuracy. We used a realistic RayTracing-based simulator with a $3$D model of a building containing outdoor UAV anchors as transmitters and indoor receivers spread across the interior of the building. By modeling Transmission, Reflection, Diffraction, and Diffused MPCs we generated a multipath channel between every receiver and transmitter. The baseline positioning performance is established using the Linear Least Squares (LLS) positioning method, which operates without any prior knowledge about the ToA measurements. Next, observe that positioning performance improves by incorporating LoS/NLoS knowledge for each ToA measurement. Further enhancement is achieved by introducing propagation awareness by recognizing that in O2I scenarios, we isolate diffraction paths by using the first arriving path principle \cite{duggal2025} and the corresponding ToA measurements can be used to generate a much improved positioning estimate using the diffraction based geometric positioning technique \cite{duggal2025diffractionaidedwirelesspositioning}. To benchmark achievable positioning performance, two different Cramér Rao Lower Bounds (CRLBs) are shown. The CRLB ($1$ Diffraction MPC) represents the lower bound on positioning accuracy assuming perfect isolation diffraction paths across all indoor node locations. The performance gap between the diffraction based geometric positioning technique and CRLB ($1$ Diffraction MPC) arises from model mismatch, as diffraction paths cannot be perfectly isolated at all indoor locations by using the first arriving principle. Finally, we have CRLB (Multiple MPC) which may provide a lower bound to propagation aware approaches. This is computed by considering all existing multipath at each indoor location and assuming perfect knowledge of the associated propagation mechanism.

\section{Concluding Remarks And Future Work}
This study explores system architectures and innovative positioning approaches designed to address signal propagation challenges in developing next-generation IPS for public safety. The  IPS relies on rapidly deployable mobile anchors that offer improved indoor signal coverage and reduced reliance on pre-installed infrastructure, taking advantage of ideas in the current scope of 3GPP standards, including NR-NTN and sidelink-related developments.
However, several ideas warrant further investigation.
\par
{\em LEO-based NTN as mobile anchors}: As NR-NTN evolves, multi-satellite-based positioning is expected to be a key feature in 6G. However, several enhancements are required to fully enable NTN-based positioning services, including but not limited to: (1) SNOs prioritizing NTN as an additional positioning service, (2) addressing satellite clock drift, as using expensive atomic clocks solely for NTN positioning is not feasible, and (3) standardizing PRS detection at UEs to enable direct NTN connectivity without relying on GNSS. Additionally, in the context of signal propagation, given the proximity of NTN in LEO orbits compared to GNSS, it is possible that the mechanism of signal propagation toward indoor locations could be via the diffraction paths instead of transmission through the building structure.

\par
{\em UAV as mobile anchors}: Since multiple UAV anchors will be deployed very close to the building for improved indoor signal coverage, this will require improved path planning and UAV swarm techniques to avoid collisions.
\par
{\em Sidelink positioning}: By enabling device-to-device positioning, some of the future directions of SLP are collaborative positioning and optimal anchor placement to maximize the coverage or visibility indoors. Although SLP is required to operate without relying on the network, synchronization between anchors is still an open problem. Future studies may explore synchronization between anchor UEs, e.g., one anchor UE connecting to the network (TN/NTN/GNSS) acting as a reference anchor synchronizing all the other anchors.  
\par
{\em Improving Propagation Aware Positioning}: Closing the gap between the CRLB positioning performance and the propagation-aware approach to positioning in \figref{fig_PU_vs_PA_plot} necessitates the development of more effective MPC classifiers. Recent approaches demonstrating significant potential have explored a combination of machine learning and ray tracing-based methods to address similar problems \cite{amiri2023indoor,hoydis2024learning}.
\vspace{-0.9em}

% \par
% Extracting propagation mechanism information from real measurements is challenging, as identifying the exact propagation mechanism responsible for a specific MPC is not straightforward. The most practical approach to addressing this challenge involves developing accurate ray-tracing-based simulators that incorporate detailed 3D models of real-world environments and employ ray-tracing algorithms to simulate signal behavior within the given environment. This could also be used to generate a large amount of training data used for the development of physics inspired machine learning techniques \cite{bakirtzis2022deepray}.

\bibliographystyle{IEEEtran}{
\bibliography{refs}
}

% \begin{IEEEbiography}[{\includegraphics[width=1in,height=1.25in,clip,keepaspectratio]{head/Gaurav_headshot.jpg}}]%
% {Gaurav Duggal} is a PhD candidate in the Wireless@VT research group at Virginia Tech.

% \end{IEEEbiography}

\end{document}

%% file: notation.tex
\def\nb0{{\mathbf{0}}}
\def\nb1{{\mathbf{1}}}

\def\figref#1{Fig.\,\ref{#1}}%

%% file: Magazine.bbl
\begin{thebibliography}{10}
\providecommand{\url}[1]{#1}
\csname url@samestyle\endcsname
\providecommand{\newblock}{\relax}
\providecommand{\bibinfo}[2]{#2}
\providecommand{\BIBentrySTDinterwordspacing}{\spaceskip=0pt\relax}
\providecommand{\BIBentryALTinterwordstretchfactor}{4}
\providecommand{\BIBentryALTinterwordspacing}{\spaceskip=\fontdimen2\font plus
\BIBentryALTinterwordstretchfactor\fontdimen3\font minus \fontdimen4\font\relax}
\providecommand{\BIBforeignlanguage}[2]{{%
\expandafter\ifx\csname l@#1\endcsname\relax
\typeout{** WARNING: IEEEtran.bst: No hyphenation pattern has been}%
\typeout{** loaded for the language `#1'. Using the pattern for}%
\typeout{** the default language instead.}%
\else
\language=\csname l@#1\endcsname
\fi
#2}}
\providecommand{\BIBdecl}{\relax}
\BIBdecl

\bibitem{li20225g}
J.~Li, K.~K. Nagalapur, E.~Stare, S.~Dwivedi, S.~A. Ashraf, P.-E. Eriksson, U.~Engstr{\"o}m, W.-H. Lee, and T.~Lohmar, ``{5G New Radio for Public Safety Mission Critical Communications},'' \emph{IEEE Commun. Stand. Mag.}, 2022.

\bibitem{firstnet2018}
J.~C. Gallagher, \emph{{The First Responder Network (FirstNet) and Next-Generation Communications for Public Safety: Issues for Congress}}, Apr. 2018.

\bibitem{9927255}
X.~Lin, ``{An Overview of 5G Advanced Evolution in 3GPP Release 18},'' \emph{IEEE Communications Standards Magazine}, 2022.

\bibitem{dureppagari_ntn_10355106}
H.~K. Dureppagari, C.~Saha, H.~S. Dhillon, and R.~M. Buehrer, ``{NTN-Based 6G Localization: Vision, Role of LEOs, and Open Problems},'' \emph{IEEE Wirel. Commun.}, 2023.

\bibitem{dureppagari2024leo}
H.~K. Dureppagari, C.~Saha, H.~Krishnamurthy, X.~F. Wang, A.~Rico-Alvari{\~n}o, R.~M. Buehrer, and H.~S. Dhillon, ``{LEO-based Positioning: Foundations, Signal Design, and Receiver Enhancements for 6G NTN},'' \emph{arXiv:2410.18301}, 2024.

\bibitem{3gpp::38845}
``{Technical Specification Group Radio Access Network; Study on scenarios and requirements of in-coverage, partial coverage, and out-of-coverage NR positioning use cases (Release 17) },'' \emph{3GPP TR 38.845, version 17.0.0}, Sep. 2021.

\bibitem{he2018design}
D.~He, B.~Ai, K.~Guan, L.~Wang, Z.~Zhong, and T.~K{\"u}rner, ``{The design and applications of high-performance ray-tracing simulation platform for 5G and beyond wireless communications: A tutorial},'' \emph{IEEE Commun. Surveys and Tutorials}, 2018.

\bibitem{amiri2023indoor}
R.~Amiri, S.~Yerramalli, T.~Yoo, M.~Hirzallah, M.~Zorgui, R.~Prakash, and X.~Zhang, ``Indoor environment learning via rf-mapping,'' \emph{IEEE Journal on Sel. Areas in Commun.}, 2023.

\bibitem{duggal2025diffractionaidedwirelesspositioning}
G.~Duggal, R.~M. Buehrer, H.~S. Dhillon, and J.~H. Reed, ``{Diffraction-Aided Wireless Positioning},'' \emph{IEEE Trans. on Wireless Commun.}, 2025.

\bibitem{duggal2025}
G.~Duggal, A.~M. Kumar, R.~M. Buehrer, N.~Kumar, and J.~H. Reed, ``{Impact of Frequency on Diffraction-Aided Wireless Positioning (To appear) preprint:arxiv},'' \emph{Proc., IEEE Intl. Conf. on Commun. (ICC)}, 2025.

\bibitem{zekavat2019handbook}
R.~Zekavat and R.~M. Buehrer, \emph{{Handbook of Position Location: Theory, Practice and Advances}}.\hskip 1em plus 0.5em minus 0.4em\relax John Wiley \& Sons, 2019.

\bibitem{3gpp_ts_38_843_r18}
``{Study on Artificial Intelligence (AI)/Machine Learning (ML) for NR air interface (Release 18)},'' \emph{3GPP TS 38.843 version 18.0.0"}, Jan. 2024.

\bibitem{guvenc2009survey}
I.~Guvenc and C.-C. Chong, ``A survey on {TOA} based wireless localization and {NLOS} mitigation techniques,'' \emph{IEEE Commun. Surveys and Tutorials}, 2009.

\bibitem{duggal20243d}
G.~Duggal, R.~M. Buehrer, H.~S. Dhillon, and J.~H. Reed, ``{3D Positioning using a New Diffraction Path Model},'' \emph{Proc., IEEE Intl. Conf. on Commun. (ICC)}, 2024.

\bibitem{hoydis2024learning}
J.~Hoydis, F.~A. Aoudia, S.~Cammerer, F.~Euchner, M.~Nimier-David, S.~Ten~Brink, and A.~Keller, ``Learning radio environments by differentiable ray tracing,'' \emph{IEEE Trans. Mach. Learn. Commun. Netw.}, 2024.

\end{thebibliography}
